# Tutoring System for Dance Learning

Rajkumar Kannan
Department of Computer Science
Bishop Heber College
Trichy 620017, TN, INDIA
rajkumarkannan@yahoo.co.in

Frederic Andres
National Institute of Informatics
University of Advanced Studies
Tokyo, JAPAN
andres@nii.ac.jp

Balakrishnan Ramadoss
Department of Computer Applns
National Institute of Tech
Trichy 620015, TN, INDIA
brama@nitt.edu

*Abstract -* **Recent advances in hardware sophistication related to graphics display, audio and video devices made available a large number of multimedia and hypermedia applications. These multimedia applications need to store and retrieve the different forms of media like text, hypertext, graphics, still images, animations, audio and video. Dance is one of the important cultural forms of a nation and dance video is one such multimedia types. Archiving and retrieving the required semantics from these dance media collections is a crucial and demanding multimedia application. This paper summarizes the difference dance video archival techniques and systems.**

## I.  INTRODUCTION

Recent advances in hardware sophistication related to graphics display, audio and video devices made available a large number of multimedia and hypermedia applications. In a scenario, one may want to digitize and store one's favorite dance presentations and later on view, say for instance, all solo dance pieces in which a particular *dancerX* was involved [16].

These multimedia applications need to store and retrieve the different forms of media like text, hypertext, graphics, still images, animations, audio and video. Multimedia database can be viewed as a controlled collection of multimedia data items. Multimedia database system, for instance dance video database, would then provide the support for multimedia data types, besides facilities for the creation, authoring, storage, access, query and control of the multimedia database.

Dance generally refers to human movements either used as a form of expression or presented in a social, spiritual or performance setting. Dance is one of the important cultural forms of a nation. The other cultural forms are art, music, and sculptures. Preserving this cultural heritage is very important for the future generations of dance community. This paper surveys the various dance video archival and retrieval systems. The organization of the paper is as follows: Section 2 deals with the dance video archival methods. The various dance analysis and retrieval systems are discussed in section 3. Section 4 deals with the challenges of Dance Analysis and Retrieval Systems (DARS). Finally section 5 concludes the paper.

## II.  DANCE VIDEO ARCHIVAL METHODS

A dance can be archived using human memories, dance notations and digital mediums. In ancient times dancers passed the knowledge of the dance verbally to the following generations. However, such knowledge was limited to the memory of the dancers and hence many dance productions had been lost due to this fact unfortunately.

*Encyclopedia Britannica* defines a dance notation as *the recording of dance movement through the use of written symbols.* Since $16^{th}$ century, dance notations are used to archive choreographies and to resurrect the dance by dancers and choreographers. A dance notation, similar to musical notes, is a symbolic form of representing the movements of the dancers using various graphical symbols such as lines, circles, rectangles, squares, bars etc.

The primary use of a dance notation is the documentation, analysis and reconstruction of choreography. Many different forms of dance notations such as *stick figure* have been created, but the two main systems used in Western culture are:

- Labanotation
- Benesh notation

Labanotation [1] is a standardized system for analyzing and recording any human movement. The original inventor is Rudolf von Laban (1879-1958), an important figure in European modern dance. He published this notation first in 1928 as *Kinetographie* in the first issue of *Schrifttanz*. Several people continued the development of the notation. In Labanotation, it is possible to record every kind of human motion. Labanotation is not connected to a singular, specific style of dance. The basis is the natural human motion, and every change from this natural human motion (e.g. turned-out legs) has to be specifically written down in the notation as shown in Figs. 1-3.

Benesh [1] movement notation invented by Benesh is particularly prominent in ballet dance and was designed to write the whereabouts of a dancer on the stage, the direction the dancer facing, the positions of the limbs, and the details of the head, hand and foot. Since dancing has movement, Benesh also notates the movement by recording the paths traced by the moving limbs. Figure 4 shows an example of Benesh notation for the human avatar shown in Fig. 3.

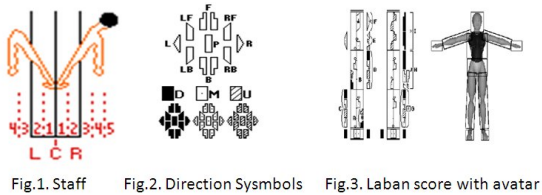
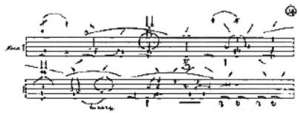

Fig.1-4. Laban and Banesh notations

A fundamental problem with Labanotation and Benesh notations is that very few people understand these notations and are experts in this field. Here the annotation process involves the expert to visualize the representation of the notation and to translate the movement to the dance learners. Further, these annotations exist in paper form only and no visual form exists.

### III. DANCE VIDEO RETRIEVAL SYSTEMS

Recording mediums [2] such as VHS and DVD can provide a quicker and cheaper means of acquiring and storing a dance. However, to capture the complete stage requires a camera to be either set back, making the dancers very small or include close-ups resulting parts of the dancers or choreography being missed or obscured. Professionally made recordings for televisions, video and DVD can capture many great productions with world-class dancers; however, the 2D camera viewpoint is subject to the director's interpretation. Also, searching these collections of digital data to learn the dance movements is tedious because of the huge volume of data.

Current research in performing arts such as dance can be broadly divided into two categories:

- Dance composition and Visualization
- Dance Analysis and Retrieval

#### A. Dance composition and Visualization

Some researches have been conducted concerning the composition of dance using the notations especially Labanotation and Benesh [3, 4, 5, 6]. Hachimura's system [7] uses markers which are placed on the body parts of the dancers and records 3D motion data. Hattori et al [8] developed key-frame based dance animation software. In his approach certain key poses in the dance are identified and are coded in Labanotation.

Several graphical editors have been developed such as LabanEditor [7], LabanWriter for Mac OS, Calaban [9] for Windows, NUNTIUS [10], LED and LINTER [11] for UNIX systems. Further, graphical editors for Benesh notation are available such as MacBenesh and Benesh Notation Editor [12]. MacBenesh is a Macintosh application that lets us create high quality single dancer Benesh movement notation scores that can be saved in a document. Benesh Notation Editor is a Windows based application for writing benesh movement notation scores. It resembles a word processor thereby the Benesh scores can be saved in a file. Several commercial softwares such as LifeForms and DanceForms [13] are also available for composing the dance interactively. These tools use virtual reality features to model the movements of the dancers.

#### B. Dance Analysis and Retrieval Systems

Dance analysis and retrieval systems perform the semantic interpretation of the dance movements in the context of culture, action, gestures and emotions. A system that directly calculates the similarities among the body motion data is described in [7] (Hachimura, 2006). It considers the issue of how a similarity of identical body motion should be defined.

Kalajdziski and Davcev [14] developed a system for the Macedonian dance videos to annotate the shots with keywords from the controlled vocabulary. The system consists of modules for segmentation, annotation, 3D dancer animation generation and Laban score generation.

Forouzan et al [15] designed a multimedia information system for the Macedonian dance annotation and the analysis of the dance features. Macedonian dance exists in the form of videos, images, audios and written commentaries. In order to process all these dance forms, the system includes several tools to extract the low level features automatically from these mediums. The visual tool extracts the visual features such as color, texture, shape, sketch, objects and their spatial relationships. The 3D motion tool extracts the motion histogram of the dancers using 3D Vicon motion capture system. This system captures data from the cameras and generates 3D motion files. Motion data are generated by interpolating the cameras' traces of translucent markers placed on the body parts of the dancers. Audio tool extracts audio features like pitch, loudness and rhythm from the audio streams. These low level multimedia features are automatically annotated by the dance experts through the annotation tool.

COSMOS-7 [16] is a multimedia annotation and retrieval system based on MPEG-7 MDS. It includes a multimedia model that integrates both low level and high level features of the video by using multimedia frames. COSMOS-7 provides a conversion tool that translates M-frames into the corresponding MPEG-7 descriptions.

Rajkumar et al [17,18,19,20] presented a semantic approach to evolve a dance video retrieval system for the Indian and generic dance videos. A central contribution of their research is the development of semantic models, annotation and authoring frameworks and query processing engines for the Indian and generic dance videos to process spatio-temporal queries. Their MPEG-7 based system allows the annotation experts to describe the metadata of the dance presentations, automatically generates the MPEG-7 instances and processes generic dance video queries. The query processor handles the dance video queries with the tree embedding technique. Also, XML based authoring and retrieval framework facilitates semantic annotation of the dance media descriptions, semi-automatic XML instance generation and query processing. The query processor allows the user to submit the knowledge based dance video queries and handles them using tree embedding technique that is supported with the domain specific ontology.

IV. CHALLENGES OF DANCE VIDEO RETRIEVAL SYSTEMS

The dance analysis and retrieval systems are best regarded as a platform that demonstrates the feasibility of choreography design, dance learning, self-paced dance learning, culture learning and dance training. However, there are a number of problems yet to be solved. The strongest qualitative drawback of these systems is the manual annotation of the dance videos, thus possibly reducing the effectiveness of DARS.

Dance retrieval models have been designed with a view of the classical and folk dance types to describe the semantics. However, there are other types of dance videos such as festival dance videos, street dance videos etc, for which some of the current semantics may not be applicable and the new semantics have to be introduced.

Semiotic data such as mood are expressed in all dance retrieval systems using the controlled vocabulary of text. Another important drawback of these systems is the absence of sound. Nevertheless, sound would enhance DARS's effectiveness involving the semiotic queries.

V. CONCLUSION

Dance generally refers to human movements either used as a form of expression or presented in a social, spiritual or performance setting. Dance video is one of the important types of multimedia resource. Since dance is the ancient cultural heritage, which is to be preserved for the future generations, archiving the dance presentations of various dance types becomes a necessary task. Further, adding a search, retrieval and faceted browsing facilities to these dance media collections by way of tools certainly provide the present and future dance trainers and learners a better eLearning environment.